\documentstyle[epsfig]{elsart}
\begin{document}
\begin{frontmatter}

\title{ Fast simulation of the Cherenkov light from showers} 
\author{L.B.~Bezrukov},
\author{A.V.~Butkevich\corauthref{cor}} 
\corauth[cor]{ Corresponding author. Tel.: 07-095-334-0188; 
Fax: +07-095-334-0184.}
\ead{butkevic@al20.inr.troitsk.ru}
\address{Institute for Nuclear Research of Russian Academy of Science,
60th October Anniversary prospect, 7a, Moscow 117312, Russia.}
\begin{abstract}
A method for fast simulation of the Cherenkov light generated by 
electromagnetic  showers is described. The parametrization for the 
longitudinal profile is used and fluctuations and correlations of the
parameters are taken into account in a consistent way. Our method 
dramatically reduces  the CPU time and its results are in rather good 
agreement with a full Monte Carlo simulation. \\
\


\end{abstract}
\end{frontmatter}
\vspace*{\fill}
\newpage
%
\section{Introduction}
\label{sect1}
\hspace*{0.6cm}
The flux of the Cherenkov photons from a shower is typically simulated 
by tracking all secondary particles of the shower down to the Cherenkov 
threshold energy $E_c$. For the electrons in water $E_c$=0.78 MeV. 
The computer time needed
for simulations of this type increases linearily with shower energy. Therefore
 a large amount of CPU time is required for simulation of a high energy
shower. However large samples of events are 
necessary to obtain results with a sufficient statistics. The parametrization
of the longitudinal Cherenkov light profile of the showers is one of the
methods to speed up simulation.\\
\hspace*{0.6cm}
A simple algorithm for parameterized showers has been successfully used for 
the simulation of the UA1 calorimeter [1]. The simulation of the longitudinal 
energy profile of electromagnetic showers was based on fitting the parameters
of a Gamma distribution to the average shower profile. Later the shape 
fluctuations of individual showers
were systematically taken into account for the simulation of calorimeter
built for the H1 experiment at HERA [2].\\
\hspace*{0.6cm}
For a fast generation of the Cherenkov light from showers with energies 
$E_0\ge$10 GeV we have developed the program FLG. The program SIMEX [3]
has been used for the full Monte Carlo (MC) simulation of the electromagnetic
showers and computation of the Cherenkov light emitted by the shower 
particles.
\section{Parametrization of electromagnetic shower}
\label{sect2}
\hspace*{0.6cm}
It is well known that the average longitudinal profile of elecromagnetic 
showers is reasonably well described by the Gamma distribution
\begin{equation}
f(t)=\frac{{\d} E}{{\d} t}=E_0 \beta\frac
{{(\beta t)}^{\alpha-1} \displaystyle exp{(-\beta t)}}{\Gamma(\alpha)},  
\end{equation}          
where $t$ is a shower depth in units of radiation length,
\begin{equation}
\int\limits_0^\infty f(t){\d} t=E_0,                        
\end{equation}
and
\begin{equation}
\frac{{\d} f(t)}{{\d} t}=\left(\frac{\alpha-1}{t}-\beta\right)f(t)       
\end{equation}          
The parameters $\beta$ and $(\alpha-1)/t$ are coefficients of absorbtion
and creation of the shower particles, respectively. There is a relation
between the location of the shower maximum $t_{\mathrm{max}}$ and the 
parameters
$\alpha$ and $\beta$,
\begin{equation}
t_{\mathrm{max}}=(\alpha-1)/\beta,                                
\end{equation}          
which follows from Eq.(1). The energy dependence of t$_{\mathrm{max}}$ can be
approximated by the expression  
\begin{equation}
t_{\mathrm{max}}=\ln{ \left(\frac{E_0}{\varepsilon} \right)} + C ,   
\end{equation} 
where $C$=-0.5 for  {\it e}-induced shower, $C$=0.5 for $\gamma$-
induced shower and $\varepsilon$=75.5 MeV is the critical 
energy for water.\\
\hspace*{0.6cm}
The longitudinal shower profile $f(t)$ reveals a useful scaling property 
when the depth of a shower is expressed in units $t_{\mathrm{max}}$ as follows:
 $\tau$=$t/t_{\mathrm{max}}$. Then Eq.(1) can be rewritten as
\begin{equation}
f(\tau)=\frac{{\d}E}{{\d}\tau}=E_0 \beta^\prime\frac
{{(\beta^\prime \tau)}^{\alpha-1} \displaystyle exp{(-\beta^\prime \tau)}}
{\Gamma(\alpha)}                                          
\end{equation}
and $\beta^\prime=t_{\mathrm{max}}\beta$. \\  
\hspace*{0.6cm}
The total distance traveled by all charged shower particles {\it L}
(track length in the units of radiation length) is proportinal to the 
shower energy $E_0$.
\begin{equation}
L \propto E_0/\varepsilon.                                  
\end{equation}
\section{Parametrization of the Cherenkov light profile of shower}
\label{sect3}
\hspace*{0.6cm}
As we deal with underwater/underice detector measuring the Cherenkov radiation
we will consider the average longitudinal Cherenkov light distribution emitted
by the shower particle only. Total number of the Cherenkov photons from the 
shower, $N^\gamma_{\mathrm{tot}}$, is proportional to the track lenght $L$. 
But according to Eq.(7) the track lenght is proportional to the shower energy,
and we can write  
\begin{equation}
N^\gamma_{\mathrm{tot}}=D_{\gamma}E_0.                              
\end{equation}          
So, the average longitudinal Cherenkov light distribution 
$f_{\mathrm{ch}}(\tau)$ can be parametrized as follows
\begin{equation}
f_{\mathrm{ch}}(\tau)=D_{\gamma}f(\tau),                                  
\end{equation}
where
\begin{equation}
\int\limits_0^\infty f_{\mathrm{ch}}(\tau){\d} \tau=N^\gamma_{\mathrm{tot}}=
D_{\gamma}E_0.                                                   
\end{equation}
Then the number of the Cherenkov photons emitted by the shower particles
in the layer $\Delta\tau$ at the depth $\tau$ is
\begin{equation}
N(\tau,\Delta\tau)=\frac{N^\gamma_{\mathrm{tot}} } {\Gamma(\alpha)}
\left [\gamma(\alpha,\beta^\prime(\tau+\Delta\tau))-\gamma(\alpha,
\beta^\prime\tau)\right],                                              
\end{equation}
where
\begin{equation}
\gamma(\alpha,x)=\int\limits_0^x t^{\alpha-1} \displaystyle{exp(-t)}{\d}t  
\end{equation}
is the incomplete Gamma function.\\
\hspace*{0.6cm}
A full MC simulation of the electromagnetic shower in water ( the radiation 
lenght $X_0$=36.1 cm) has been done using the program SIMEX. The value for the 
tracking energy threshold was set to 1 MeV. We obtain that the number of 
photons in the wavelenght 300 nm $\le \lambda \le$ 600 nm per MeV of the shower
energy is
\begin{equation}
D_\gamma=N^\gamma_{\mathrm{tot}}/E_0\approx163~{\mathrm{photons/MeV}}.    
\end{equation} 
\section{Simulation of the Cherenkov light profile taking 
 into account the fluctuations of showers}
\label{sect4}
\hspace*{0.6cm}
Realistic simulation, however, requires the simulation of individual showers.
Fluctuations of the parameters in Eq.(9) obtained from the averaged Cherenkov 
light profile does not necessarly lead to a correct description of the 
fluctuations of individual showers. Assuming that also individual shower
profiles can be approximated by the Gamma distribution
\begin{equation}
f(\tau)=E_0 \beta_i^\prime\frac
{{ \left(\beta_i^\prime \tau \right)}^{\alpha_i-1} \displaystyle 
exp{ \left(-\beta_i^\prime \tau \right)}}
{\Gamma(\alpha_i)},                                          
\end{equation}
the fluctuations and correlations can be taken into account consistently. The
 index {\it i} indicates
that the function describes an individual shower {\it i} with the
parameters $\alpha_{\it i}$ and $\beta^\prime_{\it i}$. The $\alpha_{\it i}$ 
and $\beta^\prime_{\it i}$ can be calculated from the first $\bar \tau_{\it i}$
and second $\sigma^2_{\it i}$ moments of the distribution Eq.(14) for each
single SIMEX-simulated shower:
\begin{equation}
\alpha_i=\left(\bar \tau_i/ \sigma_i \right)^2 \quad \mbox{and} \quad 
\beta^\prime_i =\bar \tau_i/ \sigma^2_i.                           
\end{equation} 
\hspace*{0.6cm}
The statistics of the full MC simulation was 1000 events for each fixed 
energy $E_0$ in the region 10 GeV $\le E_0 \le$ 10$^4$ GeV
and 200 events for $E_0$=10$^5$ GeV. The parameters $\alpha_i$ and 
$\beta^\prime_i$ are normal-distributed and such that the means 
$\langle \alpha_i \rangle$, $\langle \beta^\prime_i \rangle$ and their 
fluctuations $\sigma_{\alpha}$ and $\sigma_{\beta}$ can be determined and 
parametrized as a function of the shower energy. The average values 
$\langle \alpha_i \rangle$ and $\langle \beta^\prime_i \rangle$ vs shower 
energy are shown in Fig.1. They can be approximated by a logarithmic energy 
dependence. The correlation of the $\alpha_i$ and $\beta^\prime_i$ is given by
\begin{equation}
\rho=\frac{ \langle~(\alpha_i-\langle \alpha \rangle)
(\beta^\prime_i-\langle \beta^\prime \rangle)~\rangle }
{ [(\langle\alpha^2_i\rangle-{\langle \alpha \rangle}^2)
(\langle\beta^{\prime 2}_i\rangle - {\langle \beta^{\prime} \rangle}^2)]
^{1/2} }
\end{equation}                       
and slowly increases with shower energy from $\rho\sim0.77$ at $E_0$=10 GeV
to $\rho\sim0.85$ at $E_0$=10$^5$ GeV. \\
\hspace*{0.6cm}
In the simulation a correlated pair $(\alpha_i,\beta^\prime_i)$ is generated
to
\begin{equation}
{\alpha_i\choose \beta'_i}={\langle\alpha\rangle\choose 
\langle\beta'\rangle} +  
B{z_1\choose z_2} \qquad                     
\end{equation}
with
\begin{equation}
B=\left(\begin{array}{cc}                                      
\sigma_{\alpha}& 0\\
0& \sigma_{\beta}
\end{array}\right)+
\left(\begin{array}{cc}
\sqrt{(1+\rho)/2}& \sqrt{(1-\rho)/2}\\
\sqrt{(1+\rho)/2}& -\sqrt{(1-\rho)/2}
\end{array}\right) ,
\end{equation}
where $z_1$ and $z_2$ are standard normal-distributed random numbers.\\
\hspace*{0.6cm}
The results of the full and the fast simulations are shown in Fig.2 and 
Fig.3, respectively. In these figures the normalized means longitudinal 
Cherenkov profiles $N(\tau,\Delta \tau)/N^\gamma_{\mathrm{tot}}$ and their 
fluctuations 
$\sigma(\tau)/N(\tau,\Delta \tau)$ for {\it e}-induced showers with different
energies are given. A comparison of the SIMEX and FLG simulations shows a 
good agreement in the average profiles. The agreement in the fluctuations in 
the depth region 0.5 $\le \tau \le$ 2 improves with an increase of 
shower energy. \\
\hspace*{0.6cm}
The average angular distribution of the Cherenkov photons emitted in 
the layer $\Delta\tau$ at shower depth $\tau$ with respect to the
shower axis can be written as follows:
\begin{equation}
P(\tau,\Delta \tau,\cos \theta)=N(\tau,\Delta \tau)
\psi(\tau,\Delta \tau,\cos \theta)~{\mathrm {photons/ster}},              
\end{equation}
\begin{equation}
\mbox{where}\qquad 2\pi\int\limits_{-1}^1 \psi(\tau,\Delta \tau,\cos \theta)
{\d} \cos \theta=1                                                    
\end{equation}
The function $\psi(\tau,\Delta \tau,\cos \theta)$ was calculated for each
layer $\Delta \tau=\tau_{i+1}-\tau_i=0.1$ by the program SIMEX and use for
the fast Cherenkov light generation. The angular distribution of emitted 
photons averaged over the cascade depth,
\begin{equation}
F(\cos \theta)=\frac{1}{N^{\gamma}_{\mathrm{tot}}}
\int\limits_0^\infty P(\tau,\Delta \tau,\cos \theta){\d}\tau,        
\end{equation}
depends very slowly on the shower energy and is shown in Fig.4 vs  
$\cos \theta$ for the different values of $E_0$. In this figure 
the result obtained in Ref.[4] is also given  for comparison. The angular 
dependence of $\psi(\tau, \Delta \tau,\cos \theta)$ for different shower 
depths is shown in Fig.5. Note that the width of the 
$\psi(\tau, \Delta \tau,\cos \theta)$ distribution increases and the average 
value of $ \cos \theta $ decreases with $\tau$.\\

\section{Fast generation of the Cherenkov light from shower and 
hight energy muons}
\label{sect5}
\hspace*{0.6cm}
A procedure of the FLGf generation of the Cherenkov photons emitted by 
 a shower is the following. The shower is divided into layers with the size 
$\Delta \tau=0.1$. Each layer is considered as the point-like 
source of light. Then the flux of the Cherenkov photons from the $i$-th layer
at some point $X$ is 
\begin{equation}
\Phi_i=\left.P(\tau_i,\Delta \tau,\cos \theta)\right/R^2_i,             
\end{equation}
where $R_i$ is the distance between the center of the $i$-th layer 
and the point $X$, and $\theta$ is the angle between the shower axis and the 
direction from the center of the $i$-th layer to the point $X$. The total 
light flux at the point $X$ emitted by the entire shower is obtained as 
the sum of contributions from each shower layer. \\
\hspace*{0.6cm}
To check correctness of the simulation using the program FLG, the number
of photoelectrons collected in the PMT (Hamamatsu R2018) has been compared with
that from the full MC simulation. The detector configuration simulated for the 
comparison is given in Fig.(6). The number of the photoelectrons 
$N_{\mathrm{pe}}$
multiplied by the distance squared $R^2$ between the location of the shower 
maximum and the PMT is shown in Fig.(6) for different orientations of the PMT 
with respect to a shower with energy 1 TeV. The comparison of the FLG and 
the full MC simulation reveals that:

\begin{itemize}
\item the results are in a good agreement 
\item $N_{\mathrm{pe}}$ depends very much on the PMT direction 
with respect to the shower 
\item  at the distance $R>$20 m the shower with energy 1 TeV can be 
considered as a point-like source of light.
\end{itemize}
\hspace*{0.6cm}

The CPU time for the full simulation of the Cherenkov light is large and 
increases with the shower energy. But when we use the program FLG the CPU
time does not depend on the shower energy and decreases dramatically. For
example, at the energy $E_0$=10 GeV the CPU time for the FLG simulation 
is 3$\cdot$10$^2$ times smaller than for the full MC simulation. At the shower 
energy $E_0$=10$^5$ GeV the gain is $\approx$3$\cdot$10$^6$ times. \\
\hspace*{0.6cm}
The muon undergoes continuous (ionization) and stochastic energy loss 
processes ($\delta$-electrons, bremsstrahlung, $e$-pair production and
muon nuclear interaction). In our program for muon tracking the secondary
processes below 0.01 GeV are not simulated individually, but treated as
quasi-continuous energy loss. The additional light flux $N_a$ from these
low energy processes is parametrized by
\begin{equation}
N_a=N_0(a+b\ln{E_\mu}) .                                                
\end{equation}
Here $N_0$ is the Cherenkov light from the 'naked' muon itself with 
energy $E_{\mu}$. The full MC simulation is used for calculation of the
Cherenkov light from the secondary particles with energies 0.01$ < E <$10 GeV.
The flux of the Cherenkov photons emitted by secondaries with energies 
$E\ge$10 GeV is simulated by the program FLG.\\
\section{Conclusions} 
\label{sect6}
\hspace*{0.6cm}
A full simulation of secondary electromagnetic processes requires a large 
amount of CPU time, in particular when a low energy threshold for the
tracking ($\approx$1MeV) has to be used and when large detector volumes
are considered.
In order to speed up the simulation the program FLG has been developed.
It provides a realistic and fast simulation of the Cherenkov light from
electromagnetic showers. Shower to shower fluctuations and correlations are
taken into account consistently. The agreement between the FLG results and the
full MC is quite good. The CPU time has been dramatically reduced with the 
FLG  simulation. At the shower energy 10$^5$ GeV the gain in CPU
time is $\approx$3$\cdot$10$^6$ times. \\
\hspace*{0.6cm}
The program for the muon tracking and the FLG codes provide a fast and 
precise algorithm for large scale Monte Carlo production of high energy
events for MC study of the underwater/underice Cherenkov 
detectors like KM3.
\section{Acknowledgements} 
\hspace*{0.6cm}
We are grateful to B.Stern, S.Mikheyev and L.Dedenko for helpful 
discussions. One of us (A.B.) would like to thank A.Skasyrskaya for help
with computations.


\newpage
\section*{Figure Captions} 
Fig.1: The parameters $\langle\alpha\rangle$ and $\langle\beta^\prime\rangle$ 
of the Gamma distribution vs shower energy. The point bars are 
$\sigma_{\alpha}$ and $\sigma_{\beta}$. 

Fig.2: The normalized mean longitudinal profiles of the 
Cherenkov light $N(\tau, \Delta \tau)/N_{\mathrm{tot}}$ for {\it e}-induced 
showers with the energies E$_0=10$, 10$^3$ and 10$^5$ GeV.  

Fig.3: The relative variance $\sigma(\tau)/N(\tau,\Delta \tau)$
of the Cherenkov light profiles  for {\it e}-induced shower with energies 
E$_0=10$, 10$^3$ and 10$^5$ GeV.                 

Fig.4: The normalized integral angular distribution of the 
Cherenkov photons for different shower energies.   

Fig.5: The normalized differential angular distribution of 
the Cherenkov photons at different shower depths.   

Fig.6: $R^2*N_{\mathrm{p.e.}}$ as a function of the distance $R$ between 
the location of the shower maximum and the PMT for different directions of 
the PMT with respect to the maximum (bold star) at the shower energy 
$E$=1000 GeV. Arrows show the orientations of the PMT: V=$\cos(\theta)$.
\begin{center}
\mbox{\epsfig{file=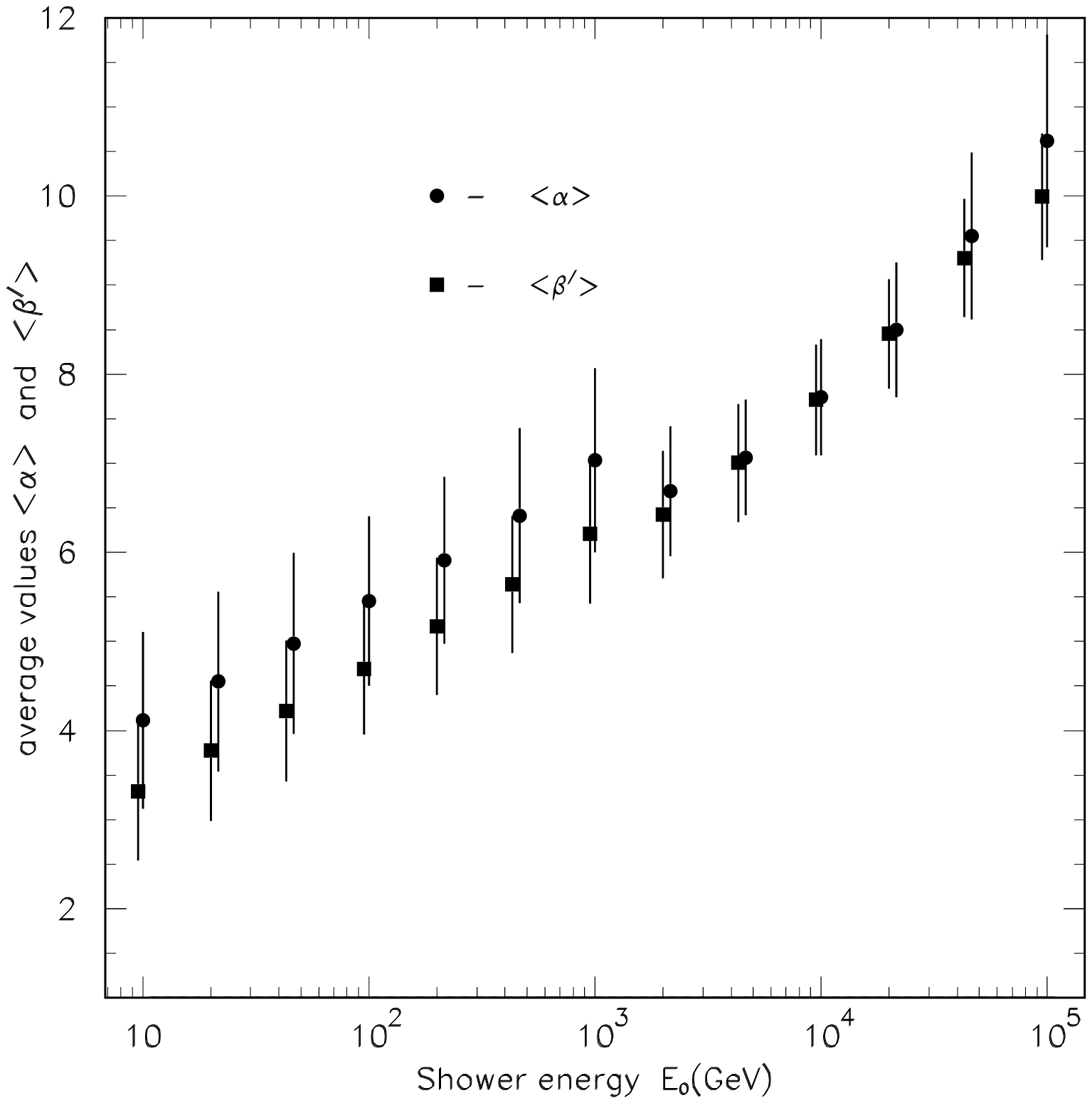,height=18cm,width=18cm}}

\vspace{2mm}
\noindent
\small
\end{center}
{\sf ~~~~~~~~~~~~~~~~~~~~~~~~~~~~~~~~~~~~~~~~~~~~~~~~~~~~~~~~~~~~~Fig.1}
\begin{center}
\mbox{\epsfig{file=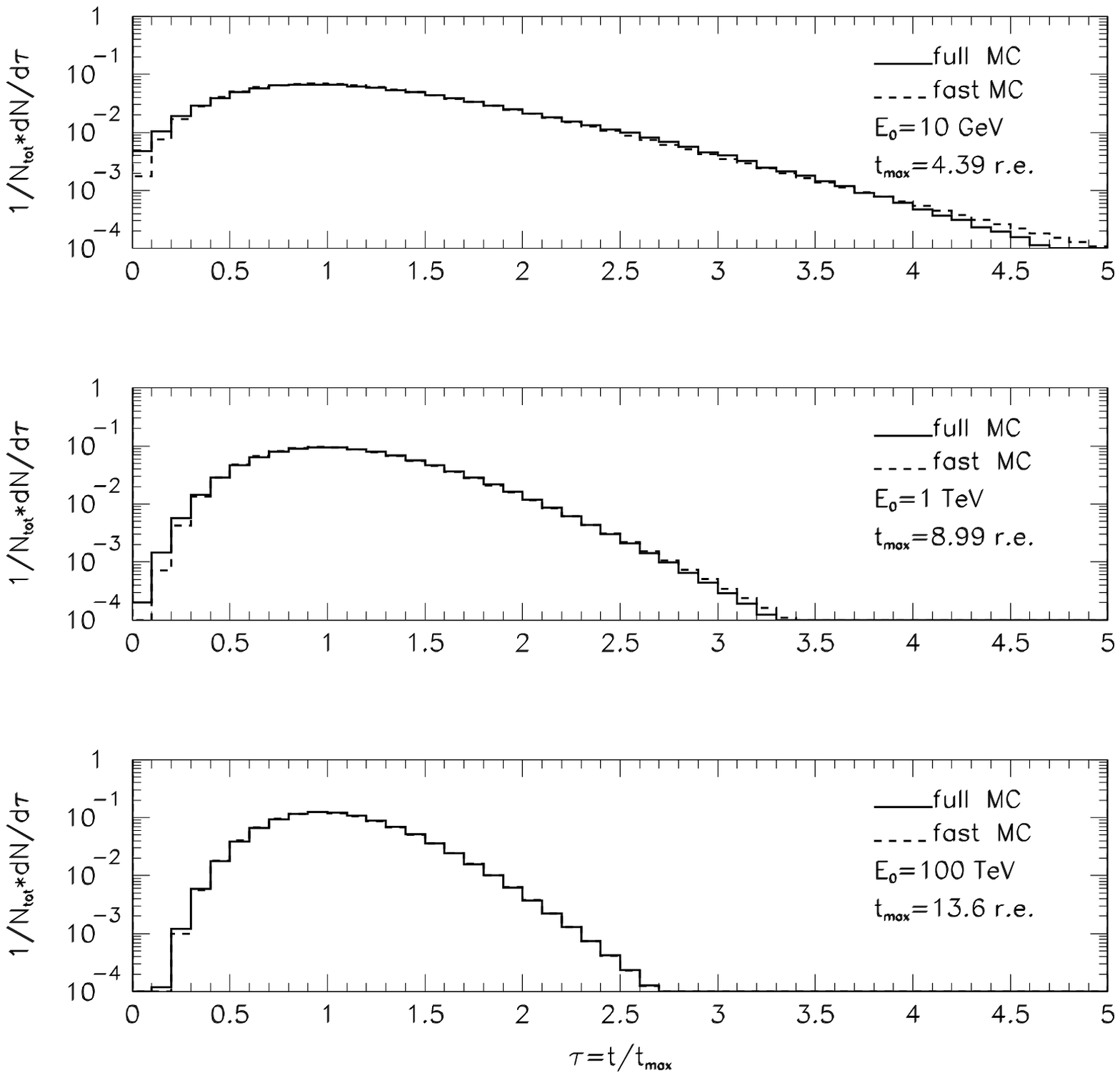,height=18cm,width=18cm}}

\vspace{2mm}
\noindent
\small
\end{center}
{\sf ~~~~~~~~~~~~~~~~~~~~~~~~~~~~~~~~~~~~~~~~~~~~~~~~~~~~~~~~~~~~~Fig.2} 
\begin{center}
\mbox{\epsfig{file=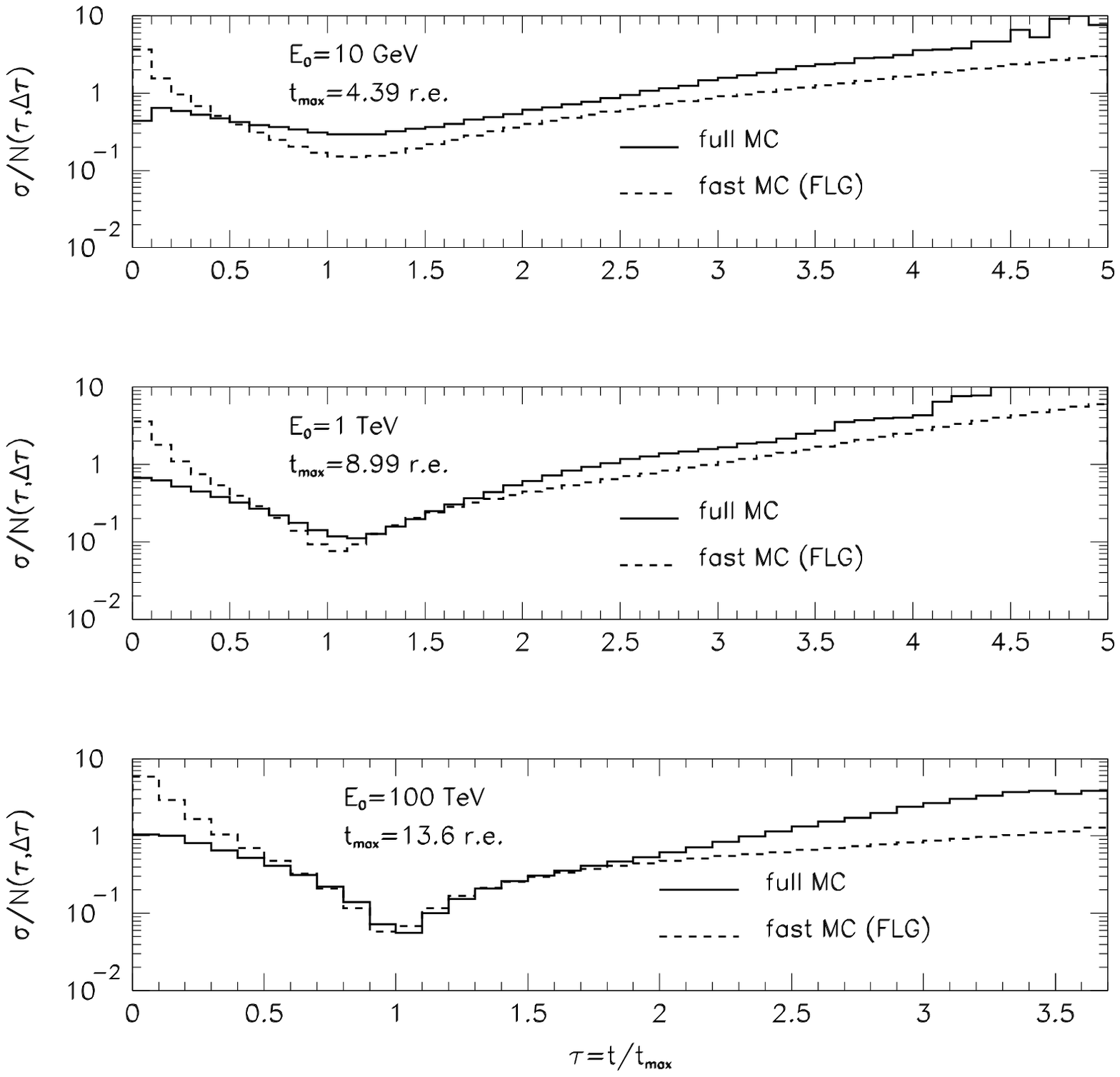,height=18cm,width=18cm}}

\vspace{2mm}
\noindent
\small
\end{center}
{\sf ~~~~~~~~~~~~~~~~~~~~~~~~~~~~~~~~~~~~~~~~~~~~~~~~~~~~~~~~~~~~~Fig.3}
\begin{center}
\mbox{\epsfig{file=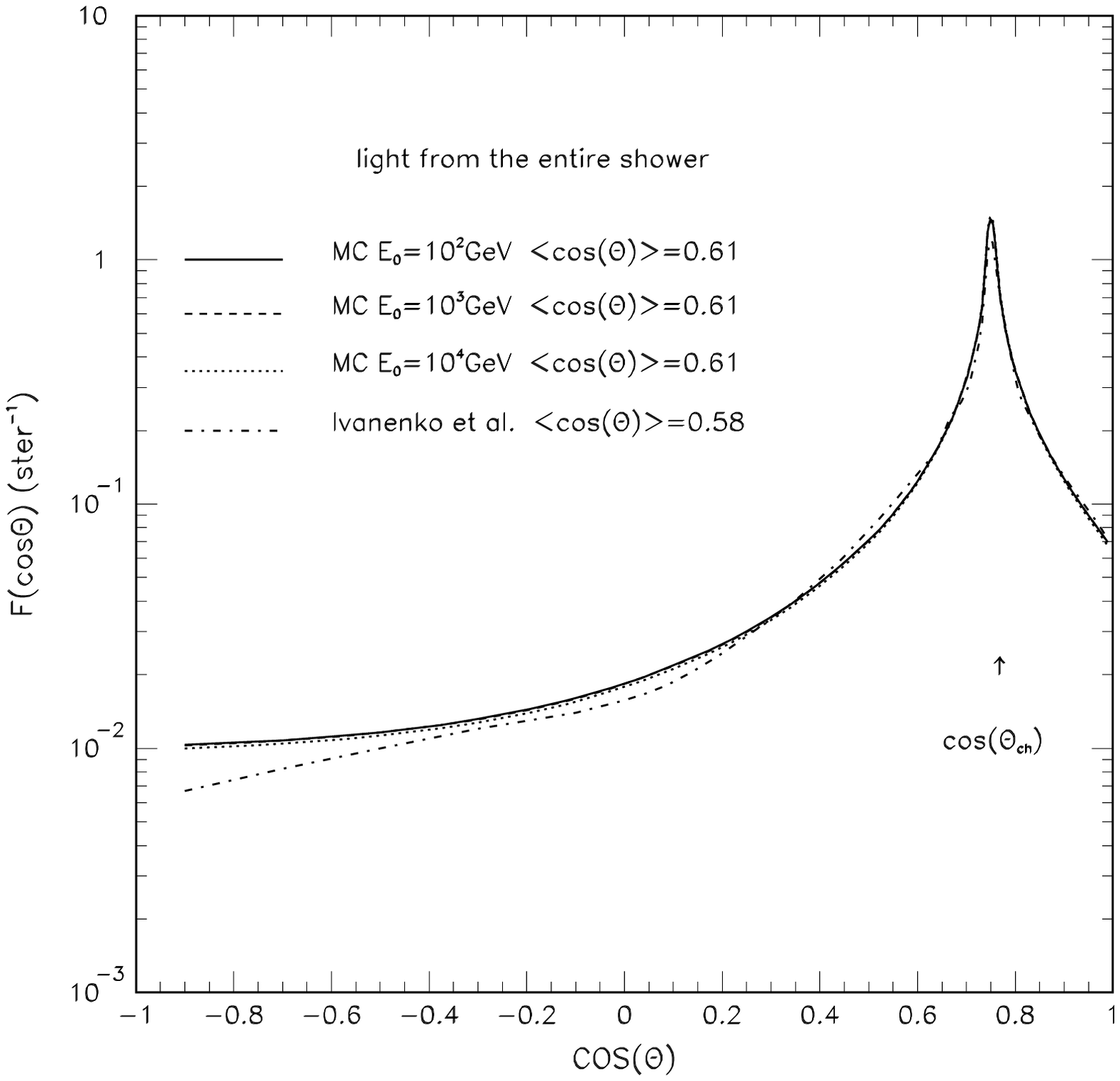,height=18cm,width=18cm}}

\vspace{2mm}
\noindent
\small
\end{center}
{\sf ~~~~~~~~~~~~~~~~~~~~~~~~~~~~~~~~~~~~~~~~~~~~~~~~~~~~~~~~~~~~~Fig.4} 
\begin{center}
\mbox{\epsfig{file=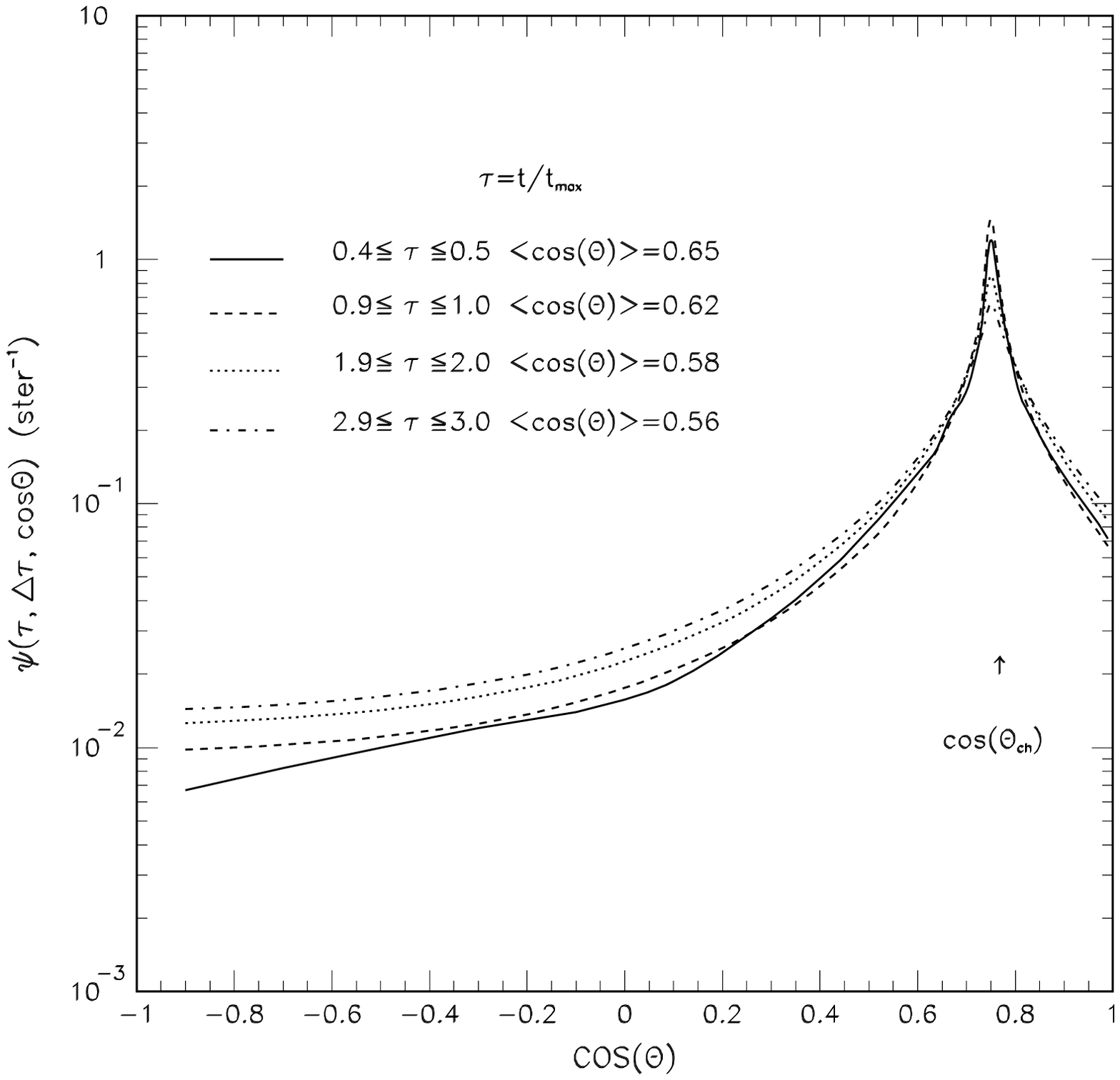,height=18cm,width=18cm}}

\vspace{2mm}
\noindent
\small
\end{center}
{\sf ~~~~~~~~~~~~~~~~~~~~~~~~~~~~~~~~~~~~~~~~~~~~~~~~~~~~~~~~~~~~~Fig.5} 
\begin{center}
\mbox{\epsfig{file=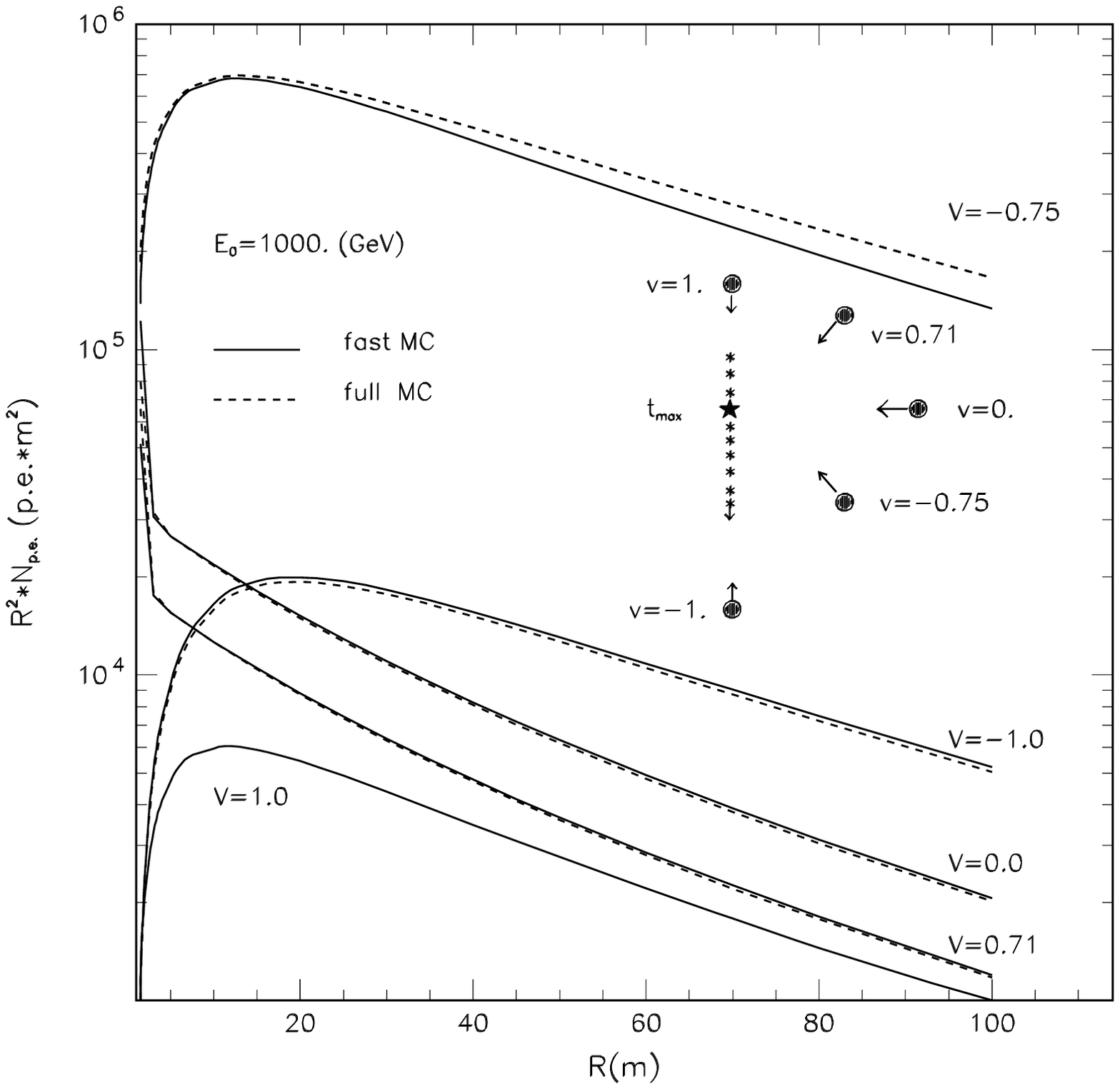,height=18cm,width=18cm}}

\vspace{2mm}
\noindent
\small
\end{center}
{\sf ~~~~~~~~~~~~~~~~~~~~~~~~~~~~~~~~~~~~~~~~~~~~~~~~~~~~~~~~~~~~~Fig.6} 
\normalsize
\end{document}